# Planar microfluidics - liquid handling without walls

A. Rathgeber, C. Strobl, H.-J. Kutschera, and A. Wixforth

*Center for NanoScience (CeNS), University of Munich, D-80539 Munich, Germany*

The miniaturization and integration of electronic circuitry has not only made the enormous increase in performance of semiconductor devices possible but also spawned a myriad of new products and applications ranging from a cellular phone to a personal computer. Similarly, the miniaturization and integration of chemical and biological processes will revolutionize life sciences. Drug design and diagnostics in the genomic era require reliable and cost effective high throughput technologies which can be integrated and allow for a massive parallelization. Microfluidics is the core technology to realize such miniaturized laboratories with feature sizes on a submillimeter scale. Here, we report on a novel microfluidic technology meeting the basic requirements for a microfluidic processor analogous to those of its electronic counterpart: Cost effective production, modular design, high speed, scalability and programmability.

Microfluidic systems miniaturize chemical and biological processes on a submillimeter scale. Reducing the dimensions of macroscopic biological or chemical laboratories is advantageous for the following reasons: The small scale allows for the integration of various processes on one chip analogous to integrated microelectronic circuitry. Thus manual handling, e.g. when transferring reagents from one process step to the next, can be reduced. Such an integration is the prerequisite for a fully automated data management system covering all steps of a given chemical or biological process. Furthermore, the required reagent volumes are reduced thus saving both material costs and process time as many of the time consuming amplification steps for biological substances can be omitted. Finally, the miniaturization results in enhanced precision by providing more homogenous reaction conditions and in shorter times for diffusion driven reactions.

Several approaches to realize microfluidic systems have been reported. One approach relies on the miniaturization of classical fluid handling devices such as pumps, valves and tubes using etching technologies known from semiconductor industry. Here, small pumping units based on e.g. piezoelectric actuation or mechanically moving parts are fabricated by deep etching processes on appropriate substrates such as glass, quartz or silicon (*1*). Closed tubes and fluid reservoirs are generally obtained by capping the etched channels with a cover plate, using anodic bonding.

Secondly, fluid handling systems have been reported that take advantage of the small dimensions of the microfluidic channel itself (*2*). As the width of a channel approaches the typical length scale (the Debye length) of the space charge regions near a solid/liquid interface, size dependent electrokinetic or electrocapillary effects can be exploited to handle and control the fluid flow. As the chemical potentials of the solid and the liquid at their interface differ considerably, a space charge region close to the interface forms. This is similar to the well-known Schottky effect in semiconductor devices. An external electric field along this space charge region induces both an electric and fluidic current close to the interface. If the channel is sufficiently narrow, this space charge current drags along the neutral fluid near the center of the channel and electro-osmotic flow is observed.

Liquid motion can also be induced by spatially modulating the wetting properties of a solid substrate. For aqueous solutions, this can be achieved by patterning the substrate such that hydrophobic and hydrophilic regions are formed. In this case, the liquid prefers to cover the hydrophilic regions and avoids to reside on the hydrophobic areas. Several techniques to realize such a modulation of the wetting properties have been reported, so far. These include microcontact printing (*3*), vapor deposition, and photolithography (*4, 5*). Fluidic motion or guided flow on the substrate can be observed (*6*) by changing the wetting properties with time. For example, a light induced guided motion of fluids has been reported using a special surface treatment, where the free energy could locally be changed under illumination (*7*).

Other research groups make use of 'peristaltic pumps' based on the ultrasonically driven deformation of thin membranes (*8*). The momentum trasfer from the membrane to the liquid induces acoustic streaming (*9*) and thus creates a guided flow along etched microchannels. Another elegant peristaltic pumphas been reported emplying polymer films (*10*) that are pneumatically deformed in the desired fashion to induce liquid flow along the channels.

So far, in most of the microfluidic systems liquids are moved in tubes or capillaries. This approach, however, suffers from the severe drawback that the pressure required to overcome surface tension and adhesion of the liquid to the walls of the channel can be enormous. As the pressure scales inversely with the minimum channel dimension, decreasing channel diameters increase the required pressure even further. Thus these systems require very powerful pumping units and valves which are not easily implemented on a microscopic scale.
Furthermore, small amounts of liquid cannot be simply handled in tubes, as these need to be completely filled in order for the pumping mechanism to work properly. The small quantities of probe liquid in this case have to be separated from each other by barriers of immiscible liquid or air bubble. In any case, however, closed tube operation is restricted to continuous flow processes. Such miniaturized systems mimic laboratories using hoses and tubes rather than beakers and eppendorfs. Most of the work in macroscopic laboratories, however, is carried out as a batch process. A typical example is the simple task of mixing two reagents or dissolving a substance in a liquid. A lab assistant would measure the required amounts in separate beakers, and then pour the substances in a third beaker while agitating with a magnetic stirrer. To date, this simple task is nearly impossible to handle given the working principles of microfluidic systems described above. Guiding liquids on a free surface overcomes the restrictions inherent to liquid handling in capillaries. However, there is no pumping mechanism available to date which would meet the requirements as defined by the lab-on-a-chip applications: Electronic control, high speed and cost effectiveness.

Here, we would like to report on a novel approach to liquid handling on smallest scales allowing for a programmable control of both batch and continous processes at high speed. Our technology meets the basic requirements of a microfluidic processor in that it is programmable and scalable, allows for massive parallelization and modular design and can be manufactured using standard semiconductor process technology.

In analogy to an electronic microprocessor, we define functional blocks, connected by fluid-conducting circuits. The confinement of the liquid to the circuits is achieved by lateral potential wells, which are defined by a controlled modification of the surface free energy (*11*). A momentum transfer from surface acoustic waves (SAW) propagating along the surface of the given substrate pumps the liquids along these fluidic circuits. Such waves are modes of elastic energy confined to a layer of about one acoustic wavelength measured from the surface of the chip. They can be excited by several means, most commonly using metallic interdigitated electrodes on top of a piezoelectric substrate. The application of a radio frequency signal to such an interdigitated transducer (IDT) leads to the excitation of a SAW with a wavelength defined by the geometry of the IDT. This small surface displacement can couple efficiently to the liquid on top of the surface and induce acoustic streaming. Typical displacement amplitudes of a SAW are in the nanometer regime, and can be externally controlled by adjusting the excitation signal. To date, SAW technology is rather advanced, as SAW devices act as radio frequency filters, resonators, and oscillators for many applications such as mobile communication and television (*12*). For the fluidic purpose as described here, the IDTs may be easily optimized in terms of actuation efficiency. For instance, single phase unidirectional transducers (SPUDTs) have been emplyoyed in our experiments. In comparison to many other liquid handling schemes, the use of SAW actuation allows for a "remote control" mode as the liquid needs not be in direct physical contact with the IDT transforming electromagnetic energy in an electromechanic excitation. The SAW rather propagates across the surface and interacts with the liquid at the desired location.

To define the fluidic trajectories on our processor we modulate the free energy of the surface laterally. For aqueous solutions, this is achieved by a lateral patterning of the wetting properties of the surface, hence defining regions which act as lateral confining potentials for the liquid. In this sense, the surface tension in microfluidic systems is no longer prohibitive but is actually exploited to define the fluidic confinement.

In Fig. 1, we show the effect of modulating the free energy of a surface to define fluidic circuits on that surface. There are different ways to achieve such a well defined modulation of the wetting properties. Here, we show our results using a homogenous silanization of the surface with subsequent photolithographic lateral patterning. To define the hydrophobic regions on the chip surface, we use octa-decyl-trichlorsilane (OTS), following a silanization procedure as described by (13). Employing this technique, a self assembled layer of about 2.6 nm thickness is formed on the surface, resulting in a wetting angle of about 110°. To define the fluidic circuits, the hydrophobic monolayer is removed using a plasma assisted ashing process, leaving a hydrophilic track behind. Meanwhile, other techniques for modulating the surface free energy such as microstructuring our samples have been tested and successfully applied. The main idea to create a lateral potential well to confine the liquid, however, remains the same in all cases.

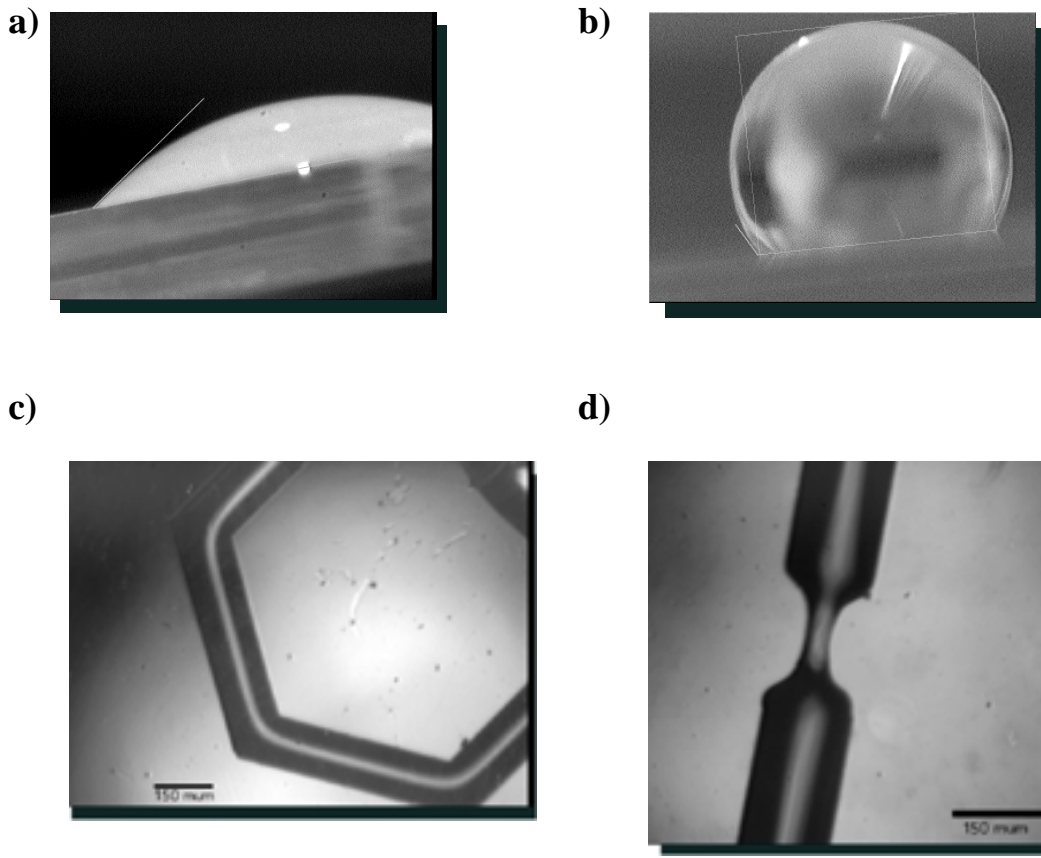

**Fig. 1:** a) Side view of a small water droplet on a hydrophilic part of the sample. The wetting angle is adjusted to about 32°, in this case. b) Side view of a small droplet on a hydrophobic part of the sample. The wetting angle in this case is increased to about 110°. c) and d) show top views of lithographically defined fluidic circuits, formed by a lateral modulation of the surface free energy.

In Fig. 1a and 1b, we show two different wetting angles on hydrophilic (a) and hydrophobic (b) areas, respectively. Fig. 1c and 1d show a top view of lithographically defined fluidic cicuits, indicating the versatility of the employed method.

As can be seen from the figure, many different geometries for such fluidic circuits and reservoirs can be realized. They can be straight or curved, they can include constrictions (see Fig. 1d) or other geometrically defined structures that may add functionality to the device. Also, the circuits do not necessarily consist of a single stripe as shown in the figure: more sophisticated, multi-striped geometries resembling railroad tracks are presently under investigation. Small amounts of liquid are deposited on the processor surface using a conventional, piezoelectrically driven ink jet nozzle, directed towards a hydrophilic region of the chip serving as a reservoir for the fluid. Shallow etching of these compartments may be used to enhance the confinement. Quantities as small as one nanoliter and below can be applied to the chip in an accurate and reproducible manner.

The whole microfluidic processor is encapsulated to prevent the chip from contamination and to control the thermodynamics, for instance to maintain a given relative humidity and to avoid the evaporation of small droplets. For specific applications, it is possible to cool or heat parts of the microfluidic processor, e.g. using meander-like on-chip heaters or a Peltier element. The package also contains the ink jet nozzle(s) to 'load' specific reservoirs for further acoustically driven dispensing on the chip.

In our case the chip is either made of $LiNbO_3$ or quartz ($SiO_2$). Both materials are standard piezoelectric substrates for SAW applications. To achieve biocompatibility the $LiNbO_3$–chip is usually covered with a thin (0.1μm to 3μm) $SiO_2$ layer, using a plasma enhanced chemical vapor deposition (PECVD) process. For specific applications, however, the surface of the substrate may be covered by a thin film of almost any desired material, including metals such as gold. The surface acoustic waves are excited by metallic interdigitated transducers (IDT) of different application-related designs, including single phase unidirectional transducers (SPUDT) to increase the pumping efficiency. The IDT are deposited by a standard planar process on appropriate sites of the chip. If the transducers are fed with a radio signal frequency (in the present case 150 MHz on a $LiNbO_3$-chip), a SAW with a wavelength of about 23 μm is launched along the desired direction. Both continous wave and pulsed excitation of the SAW are possible, depending on the application. Other substrates may also be used, as long as they support surface acoustic waves, which most crystalline materials do. The substrates need not be piezoelectric themselves. It is sufficient to deposit a piezoelectric material such as ZnO on the regions of the surface, where the IDT's are located.

The interaction between the SAW and the liquid on the chip surface leads to a momentum transfer and the application of a net force from the wave to the liquid. Depending on the amplitude of the SAW and the confinement of the fluid this force can either induce acoustic streaming inside the liquid or it can move the whole volume across the surface of the chip. The acoustic streaming within the liquid can be employed to pump the entire volume without moving its center of mass and/or to actuate suspended particles in the liquid. Also, turbulences within the liquid can be excited which effectively stir liquids on smallest scales. Once the amplitude of the SAW exceeds a critical value, the pinning of the boundary is overcome and the whole volume is actuated, i.e. its center of mass changes its position. Here, the liquid can be moved either in form of single droplets or in larger volumes on a fluidic circuit. This critical SAW amplitude depends on the actual volume of the droplet, the specific geometry of the fluidic circuit, and the surface properties which can be easily understood in terms of well known pinning forces at the three-phase boundary between the liquid, the solid surface and the surrounding gas. In any case, however, the critical SAW amplitude and the required pulse sequence can be determined very reproducibly. Hence, the mode of operation (streaming, stirring, pumping, actuation of droplets or a combination thereof) can be predetermined and is electrically controllable. The laterally patterned fluidic circuits keep the liquids confined to the tracks as long as the SAW amplitude (or the equivalent effective pressure) does not exceed a second critical value. This property of the system can be used, for instance, to rinse the whole structure with an appropriate liquid.

In Fig. 2, we show a time series for an acoustically driven small droplet (V≈ 50 nl) being guided along a fluidic circuit of about 300 μm width. A series of SAW pulses of about 1 msec length at a repetition rate of 10 msec propagates from left to right, 'pushing' the droplet along the track. The velocity of the droplet can be adjusted by the amplitude of the SAW, the pulse width and/or the repetition rate. In the present experiment, a rather slow movement of the droplet (about 2.35 mm/sec) has been chosen. It should be noted, however, that very high droplet velocities up to 5 cm/sec have been observed, so far. This limit, however, is presently given by the time resolution of our experimental setup.

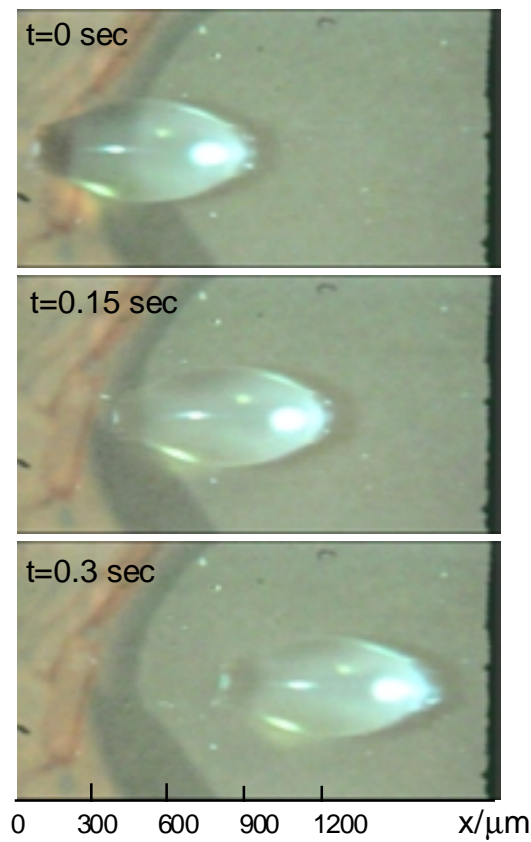

**Fig. 2:** Time series of an acoustically actuated single droplet (approx. 50 nl), moving along a fluidic track defined by a controlled lateral modulation of the surface free energy. The speed of the droplet can be adjusted by the amplitude or the timing sequence of the pulsed surface acoustic wave. In the present experiment, the speed is set to about 2.35 mm/sec, which is rather slow for our technique. The droplet can be moved over large distances and then be brought to a stop at a desired position on the chip. Note the nice side effect that can be seen from the figure: The possibility to realize a moveable and adjustable microlens for optical readout or excitation purposes.

Such a moving droplet is very well suited to deliver smallest amounts of liquid to certain areas on the chip, thus 'loading' the spots of, for instance a microarray, with a desired chemical or fluid. For this purpose, we have fabricated a device with small areas of reduced wetting angle arranged in a checkerboard geometry. The geometry, the proper design of the surface chemistry and the speed of the droplet define the size and the volume of the deposited droplets, in our case less than 50 picoliters, each.

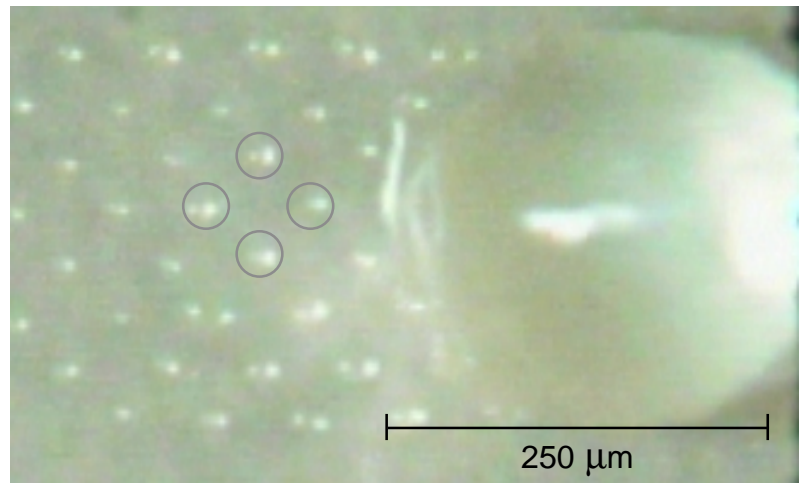

**Fig. 3:** Loading of a microarray with a spot distance of about 70 µm using an acoustically actuated droplet moving from left to right acting as a reservoir. Within the fluidic track, a checkerboard geometry of specially treated surface areas has been lithographically defined. The large droplet of about 50 nl volume selectively wets these areas, leaving an array of small droplets behind, each having a volume of only 50 picoliters. The small circles have been added to the figure to enhance and visualize the contours of the smaller droplets. Again, the droplets and the droplet array can be used as a moveable and adjustable optical system of lenses for readout purposes.

Moreover, many droplets on different circuits can be moved independently of each other, employing different SAWs with different propagation directions and/or amplitudes. Using two counterpropagating SAWs it is possible to move two droplets towards each other. Here, we make use of the fact that the SAW is strongly attenuated by a liquid in its propagation path. The amplitude of the SAW moving from, say left to right, is strongly reduced behind the droplet being pushed from left to right. Hence, this SAW will have a significantly reduced effect on the droplet moved by the second SAW in the other direction.

Using such 'printed circuits', programmable fluidic processors with decoupled functional blocks can be designed to perform many different tasks. Not only are we able to control the trajectory of the droplet with high accuracy. By using different SAWs and different droplets, a typical laboratory batch process can be designed. For instance, two droplets of different fluids can be moved independently of each other towards a 'reaction chamber', where they are merged and - again acoustically driven - mixed in a controlled manner to significantly speed up the reaction for the required chemical or biological processes. In many conventional microfluidic devices, the reaction speed is generally limited by diffusion as small amounts of liquid often exhibit laminar rather than turbulent flow, depending on the relevant Raynolds number. The slowness of the diffusion is a severe draw-back for current lab-on-a-chip applications since typical areas of e.g. DNA microarrays are as large as 1 cm$^2$ and above.

To demonstrate the power of the acoustically driven microfluidics described here, we show in Fig. 4 three snapshots of a chemical reaction between two small droplets. The volume is approximately about 40 nanoliters each. We chose a chemoluminescence experiment, based on the luminole reaction that is routinely used as a marker in diagnostics. Both droplets are acoustically pumped towards each other until they merge and react. The same SAWs that actuate the droplets are then used to stir the compound and hence provides more homogenous reaction conditions within the sample. Within the time resolution of our experiment, given by the frame rate of our CCD camera of about 20msec, the reaction occurs nearly instantaneously across the whole volume of the combined droplets. This can be nicely traced by observing the luminescence in the liquid. No special care has been taken to control the mixing process beyond the acoustical stirring. In principle, the geometry of the 'reaction' chamber could be adjusted such that the mixing occurs even more homogeneously as in the experiment shown here.

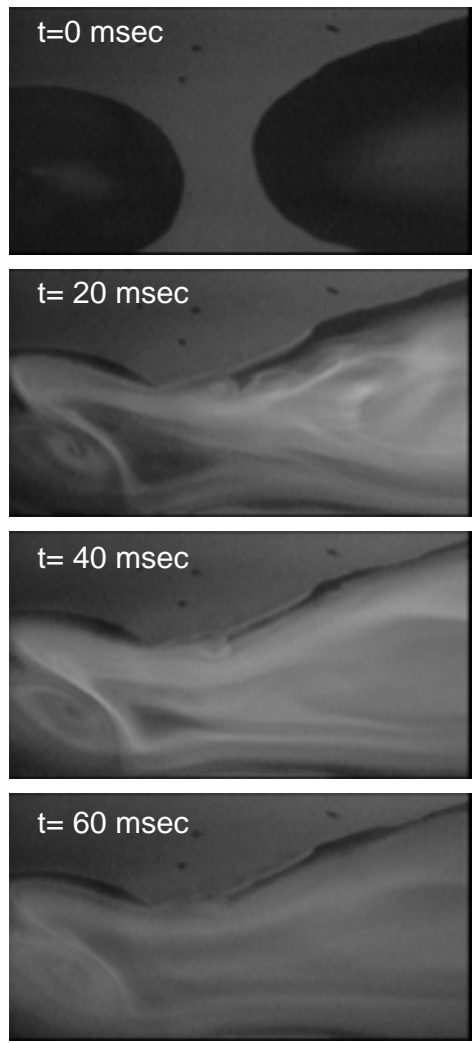

**Fig. 4:** An acoustically induced chemical reaction between two small droplets. Both droplets are moved independently towards each other until they merge. The acoustic pumping induces strong turbulences within the combined droplets, thus dramatically increasing the speed of the reaction.

So far, we have only regarded the purely mechanical interaction between a SAW and (fluid) matter on the surface of the substrate. However, as SAWs on piezoelectric substrates are accompanied by strong lateral and vertical electric fields, a direct coupling of these fields to polarizable or charged substances is also possible. Such an acoustoelectric coupling has been used in the recent past to study the interaction between SAWs and mobile electrons and holes in semiconductor layered systems (*14*) to acoustically move those charges along the plane of a semiconductor quantum well (*15*). For our microfluidic processor, direct acoustoelectric coupling can also be exploited for charged or polarizable media. For instance, the superposition of two or more SAWs leads to an interference pattern with well defined nodes of zero electric field and zero mechanical deformation of the surface. This can be used for both an additional electric confinement, similar to acoustic levitation in intense sound fields (*16*), and for transporting media by moving nodes of zero electrical and/or mechanical potential across the surface. Hence, the electrically driven actuation of liquids and particles, molecules and cells suspended in a liquid is also possible with planar fluidics.

Another physical effect, occurring at free surfaces can be exploited to realize pressure dependent valves or switches, which are essential parts of a microfluidic processor. Neglecting the effects of gravity, the pressure within a small amount of liquid is constant. This constant pressure determines the shape of the droplet and leads to the fact that the curvature of thze droplet surface is constant (exept for the circumference and the parts touching the surface). Therefore, a constriction (similar to the one shown in Fig. 1d) can act as a virtual barrier within a fluidic circuit. However, if a pressure (or momentum) above a certain threshold is applied to the liquid it can overcome the constriction without leaving the hydrophilic area. Meanwhile, we have successfully demonstrated the functionality of such pressure dependent valves and switches and have used this phenomenon to separate small amounts of liquid from a bigger volume serving as a reservoir. The effect can also be used to measure the amount of liquid within a small compartment, a technique that will be reported elsewhere.

It should be noted at this point, that the very same SAW technology described here can also be used for sensor applications (*17*) on the same chip used for liquid actuation. Hence, we are able to add even more functionality to our microfluidic processor. SAW sensors are already used for many different applications, mostly based on the sensitivity of SAW's to loading a free surface with a finite mass. A simple sensor application for our fluidic processor would for instance consist of a feedback system with one SAW moving a droplet from one location to another, while the transmission of a second SAW is used to determine whether the droplet has arrived at its designated position. Here, we make use of the fact that liquids on a surface strongly attenuate SAWs. Using a specially designed IDT structure allowing for a spatial resolution, even more sophisticated sensors can be designed (*18*). We use so-called slanted transducers which transmit a relatively narrow SAW-beam with the specific point of excitation and sound path defined by the actual SAW frequency. This narrow SAW beam can be swept along the whole aperture of the slanted transducer by modulating the excitation frequency. The transmitted SAW intensity can then be used as the sensing quantity. Also, a two-dimensional spatial resolution is possible using two sets of slanted transducers with two perpendicularly propagating SAWs. The principle of such two-dimensional SAW sensors has been shown before to be sensitive to the presence of free charges close to the surface or mass loading effects (*18*). Of course, such slanted transducers can also be employed for pumping and stirring liquids. Here, the spatial tunability of the sound path gives additional flexibility for, e.g., multiplexing different fluidic circuits.

In summary, we have realized the basic features of a programmable microfluidic processor by combining the lateral modulation of the surface free energy with a SAW driven liquid handling system. We exploit the increased contribution of surface tension effects to confine the liquids under investigation to defined fluidic circuits for guided flow and to realize pressure dependent valves and switches. The independent and controlled movement of different droplets is feasible employing different acoustic waves which can be controlled separately. The interaction between the surface waves and the liquid leads to streaming effects and turbulences, which can be used to induce and accelerate physical, chemical or biological reactions or to solve solid substances in a liquid. As our technology allows for the movement of single droplets across the surface different functional blocks are effectively decoupled making it possible to describe the whole system as a sum of its parts. Furthermore, our acoustically driven liquid handling system is scalable and the pumping mechanism has been found to work for volumes as small as several picoliters all the way up to several 100 µl. These are the prerequisites to build design libraries similar to those used in the semiconductor industry thus reducing the development time for new and complex systems. Our experimental results suggest that the acoustically driven liquid handling on a free surface has the potential to define a new generation of microfluidic processors that comprise all the characteristics of their electronic counterparts.

We gratefully acknowledge very useful discussions with J. P. Kotthaus, and Th. Bein (Center for NanoScience, Munich), P. Leiderer, and C. Schäfle (University of Konstanz), and J. Scriba, E. Neuhaus, and Ch. Gauer (Advalytix AG). This work has been financially supported by Advalytix AG.